\documentclass[12pt,reprint,aps,groupedaddress,showpacs,nofootinbib,prd,twocolumn]{revtex4-2}

\usepackage{amsmath,amssymb,graphicx,amsfonts}
\usepackage{mathrsfs}
\usepackage{comment}
\usepackage{xcolor}
\usepackage[shortlabels]{enumitem}
\usepackage{hyperref}
\usepackage{url}

\newcommand{\beq}{\begin{equation}}
	\newcommand{\eeq}{\end{equation}}

\newcommand{\bse}{\begin{subequations}}
	\newcommand{\ese}{\end{subequations}}

\newcommand{\ain}[1]{$\boldsymbol{#1}$}
\newcommand{\ainF}[1]{\boldsymbol{#1}}

\begin{document}
\title{Gravitational-wave imprints of compact and galactic-scale environments in extreme-mass-ratio binaries}

\author{Kyriakos Destounis$^{1,2,3}$, Arun Kulathingal$^3$, Kostas D. Kokkotas$^{3,4}$ and Georgios O. Papadopoulos$^4$} 
\affiliation{$^1$Dipartimento di Fisica, Sapienza Università di Roma, Piazzale Aldo Moro 5, 00185, Roma, Italy}
\affiliation{$^2$INFN, Sezione di Roma, Piazzale Aldo Moro 2, 00185, Roma, Italy}
\affiliation{$^3$Theoretical Astrophysics, IAAT, University of T{\"u}bingen, 72076 T{\"u}bingen, Germany}
\affiliation{$^4$Section of Astrophysics, Astronomy, and Mechanics, Department of Physics, University of Athens, Panepistimiopolis Zografos GR15783, Athens, Greece}

\begin{abstract}
Circumambient and galactic-scale environments are intermittently present around black holes, especially those residing in active galactic nuclei. As supermassive black holes impart energy on their host galaxy, so the galactic environment affects the geodesic dynamics of solar-mass objects around supermassive black holes and subsequently the gravitational waves emitted from such non-vacuum extreme-mass-ratio binaries. Only recently an exact general-relativistic solution has been found that describes a Schwarzschild black hole immersed in a dark matter halo profile of the Hernquist type. We perform an extensive analysis generic geodesics delving in such non-vacuum spacetimes and compare our results with those obtained in vacuum Schwarzschild spacetime, as well as calculate their dominant gravitational-wave emission. Our findings indicate that the radial and polar oscillation frequency ratios, which designate resonances, descend deeper into the extreme gravity regime as the compactness of the halo increases. This translates to a gravitational redshift of non-vacuum geodesics and their resulting waveforms with respect to the vacuum ones; a phenomenon that has also been observed for ringdown signals in these setups. We calculate the maximized overlap between waveforms resulting from orbital evolutions around Schwarzschild and non-vacuum geometries and find that it decreases as the halo compactness grows, meaning that dark matter environments should be distinguishable by space-borne gravitational-wave detectors. For compact environments, we find that the apsidal precession of orbits is strongly affected due to the gravitational pull of dark matter; the orbit's axis can rotate in the opposite direction as that of the orbital motion, leading to a retrograde precession drift that depends on the halo's mass, as opposed to the typical prograde precession transpiring in vacuum and galactic-scale environments. Gravitational waves in retrograde-to-prograde orbital alterations demonstrate transient frequency phenomena around a critical non-precessing turning point, thus they may serve as a `smoking gun' for the presence of dense dark matter environments around supermassive black holes.
\end{abstract}
	
\maketitle

\section{Introduction}

One of the most curious and enigmatic conundrums that puzzles the physics community for decades is the dark matter problem in our Universe. Although we are now convinced that the striking majority of the mass in the Universe is indeed comprised of non-ordinary (and non-luminous) matter, there are still efforts to understand its composition \cite{Bertone:2004pz,Clowe:2006eq,Bergstrom:2009ib}. The current Standard Model of cosmology, namely the $\Lambda$ Cold Dark Matter ($\Lambda$CDM) model, describes the structure formation of the Universe, from stars to galaxy clusters, quite successfully in accord with the observational characteristics of the Universe \cite{DelPopolo:2007dna,Primack:1997av}, though there are still a number of challenges to be resolved \cite{Perivolaropoulos:2021jda}. 

Even though dark matter is not directly perceptible \cite{Freese:2008cz,Kahlhoefer:2017dnp}, there is a plethora of indirect evidence for the existence of a field that only interacts gravitationally \cite{Munoz:2003gx,PerezdelosHeros:2020qyt}. Some examples are the discrepancy in rotation curves of galaxies and the fact that they cannot be explained only by ordinary matter \cite{Oort:1940,Rubin:1970,Rubin:1980,Begeman:1991iy,Persic:1995ru,Corbelli:1999af}, the inconsistency between gravitational lensing predicted by General Relativity (GR) and observations \cite{Moustakas:2009na,Massey:2010hh,Ellis:2010}, as well as the observed Cosmic Microwave Background Radiation power spectrum that strongly supports the existence of dark matter \cite{Challinor_2012}, to name a few.

Since dark matter is assumed to interact only gravitationally, it should be taken into account in gravitational-wave (GW) astrophysics \cite{Barack:2018yly}, especially because astrophysical environments are omnipresent in galactic media. In the meantime, the first GW detection of a black hole (BH) binary merger by the LIGO/Virgo Collaboration \cite{LIGOScientific:2016aoc}, and subsequent ones \cite{LIGOScientific:2021djp}, has opened an entirely new avenue for precision GW astronomy. GWs carry pristine information regarding the binary's constituents, as well as the final remnant's externally observable quantities. So far, the majority of mergers observed last for fractions of a second. Although GW astronomy is in full bloom and we are now able to extract significant information regarding the spacetime geometry and the Kerrness of BHs, it is natural to venture out and explore novel GW sources that are at the moment inaccessible with current detectors.

The Laser Interferometer Space Antenna (LISA) \cite{LISA:2017pwj} is a space-borne GW detector that will open new realms in GW astrophysics, due to its unprecedented level of accuracy, and pursue in particular mHz sources of GWs \cite{Baibhav:2019rsa,Amaro-Seoane:2022rxf,LISA:2022kgy,Karnesis:2022vdp}. One of the prime objectives of LISA (and other space programs \cite{TianQin:2015yph,Ruan:2018tsw,Ruan:2020smc}) is the detection of gravitational radiation from extreme-mass-ratio inspirals (EMRIs) \cite{Gair:2017ynp}, which comprise of a \emph{primary} supermassive BH and a \emph{secondary} stellar-mass compact object. Supermassive BHs, and consequently EMRIs, reside in stellar clusters and galactic cores, thus including environmental effects in waveform models should be prioritized in order to maximize the science yield of space-based detectors \cite{Barausse:2006vt,Barausse:2007dy,Eda:2013gg,Macedo:2013qea,Barausse:2014tra, Cardoso:2016olt,Cardoso:2019rou,Kavanagh:2020cfn,Toubiana:2020drf,Caputo:2020irr,Traykova:2021dua,Zwick:2021dlg,Zwick:2022dih,Speri:2022upm,Sberna:2022qbn,Polcar:2022bwv,Vicente:2022ivh,Speeney:2022ryg}.

Even so, the bulk of EMRI analyses treat these systems in vacuum, or with Newtonian potentials that approximate matter and dynamical friction. Only recently an exact general-relativistic solution of the Einstein field equations has been obtained \cite{Cardoso:2021wlq}, that describes a BH immersed into a dark matter halo of the Hernquist type \cite{Hernquist:1990be}, and was further extended to different dark matter profiles \cite{Konoplya:2021ube,Stuchlik:2021gwg,Jusufi:2022jxu,Konoplya:2022hbl}. The axial and polar GW fluxes have been recently investigated in a generalized setup \cite{Cardoso:2022whc}, for circular equatorial EMRIs, and strongly support the need for astrophysical environments to be taken into consideration in EMRI waveform modeling.

In this study, we investigate the orbital phase space of generic, non-circular and precessing geodesics in galactic-scale and compact dark matter environments, as well as their emitted GW radiation, without restricting the orbit onto the equatorial plane. We find that geodesics are integrable irregardless of the halo's compactness and that bound orbits occupy a larger volume in phase space with respect to Schwarzschild geodesics when the secondary's properties remain fixed. We further find that the characteristics of the halos considered impose significant changes in phase space and lead to orbits with longer revolution period, due to dynamical friction, that possess larger orbital frequency ratios as the halo compactness increases. This translates to a significant decrease in the match between vacuum Schwarzschild and non-vacuum GWs from such asymmetric binaries, even when the compactness is of order $10^{-6}$ and can represent a galactic-scale environment, as well as a redshift in their respective GW frequencies.
	
Taking the compactness of the halo at a large limit, where the solution still does not violate any energy conditions or exhibits external singularities, we find a clear phenomenological imprint, where the trajectories can experience a retrograde-to-prograde precession transition and the GW frequencies slowly convert from triplets to single Fourier peaks and back as the secondary crosses a critical radial position. This phenomenon only occurs when the dark matter halo is compact and massive enough; in such case the dark matter pull can antagonize the general-relativistic effects of the primary. Our analysis provides results both in the orbital and waveform level, and assesses potential effects of astrophysical and compact environments during an inspiral's progression under the assumption of adiabatic evolution through a successive geodesic scheme. In what follows we utilize geometrized units so that $G=c=1$.

\section{Black holes in galaxies}

We operate on an exact solution of Einstein's equations that describes a non-rotating BH lurking in the center of a galactic dark matter halo \cite{Cardoso:2021wlq}. The construction assumes many gravitating masses following all possible geodesics and surrounding the central object, thus building an Einstein cluster. The construction is equivalent to introducing an anisotropic material with vanishing radial and non-vanishing tangential pressure $P_t$, such that
\begin{equation}
	T^\mu_\nu=\text{diag}(-\rho,0,P_t,P_t),
\end{equation}
where $\rho$ describes the dark matter profile. Even though there are plenty of density profiles to describe dark matter halos \cite{King:1962wi,Jaffe,Navarro:1995iw,Zhao:1995cp}, an exact spacetime geometry has only been found when the Hernquist density profile is utilized, namely \cite{Hernquist:1990be}
\begin{equation}\label{Hernquist}
	\rho=\frac{M a_0}{2\pi r (r+a_0)^3},
\end{equation}
where $M$ is the mass of the halo, $a_0$ its length scale and $M/a_0$ defines the halo compactness. The assumption of spherical symmetry, together with a Hernquist-inspired matter distribution
\begin{equation}
	m(r)=M_\text{BH}+\frac{M r^2}{(a_0+r)^2}\left(1-\frac{2M_\text{BH}}{r}\right)^2,
\end{equation}
where $M_\text{BH}$ the mass of the primary BH, leads to the spacetime geometry
\begin{equation}\label{metric}
	ds^2=-f(r)dt^2+\frac{dr^2}{1-2m(r)/r}+r^2 d\Omega^2,
\end{equation}
with 
\begin{align}
	f(r)&=\left(1-\frac{2 M_\text{BH}}{r}\right)e^\Upsilon,\\
	\Upsilon&=-\pi\sqrt{M/\xi}+2\sqrt{M/\xi}\arctan\left[\frac{r+a_0-M}{\sqrt{M\xi}}\right],\\
	\xi&=2a_0-M+4 M_\text{BH}.
\end{align}
At small scales, Eq. \eqref{metric} describes a BH of mass $M_\text{BH}$, while at large distances the Newtonian potential corresponds to that of the Hernquist profile \eqref{Hernquist}, dominated by $M$. The causal structure of spacetime consists of an event horizon at $r=2 M_\text{BH}$, a curvature singularity at $r=0$, while the configurations has Arnowitt-Deser-Misner (ADM) mass equal to $M+M_\text{BH}$. For astrophysical scenarios, such as galactic-scale halos, the inequality $M_\text{BH}\ll M\ll a_0$ should hold together with compactness of order $M/a_0\lesssim 10^{-4}$ \cite{Navarro:1995iw}. Nevertheless, in the context of BH environments, the compactness is a free parameter as long as $M< 2(a_0+2M_\text{BH})$, in order to avoid further curvature singularities besides the one at $r=0$. In the rest of this analysis, we will conform to the aforementioned inequality.

\section{Orbital dynamics}

The most suitable way to adiabatically evolve an EMRI is through the calculation of the axial and polar GW fluxes in order to drive the inspiral through successively damped geodesics. Currently, the most proper EMRI analysis in the geometry \eqref{metric} has been performed in \cite{Cardoso:2021wlq,Cardoso:2022whc} for circular equatorial orbits. Nevertheless, a first-order approximation to EMRI evolution can be accomplished through geodesics of a test-particle which plays the role of the secondary orbiting around the primary supermassive BH. As such, one can gain important intuition regarding the elemental structure of the underlying background geometry at the geodesic level. {In what follows, we will consider generic orbits, that even though are planar due to the spherically-symmetric nature of the primary, are not circular and are precessing, therefore the initial conditions and parameters of the secondary are not fine-tuned but rather satisfy appropriate constraints for bound geodesic motion.

\subsection{Geodesic evolution}

The geodesic equations read
\begin{equation}\label{geodesic}
	\ddot{x}^\kappa+\Gamma^\kappa_{\lambda\nu}\dot{x}^\lambda \dot{x}^\nu=0,
\end{equation}
where $\Gamma^\kappa_{\lambda\nu}$ are the Christoffel symbols associated with the background spacetime, $x^\kappa$ is the four-position, $\dot{x}^\kappa$ is the four-velocity and the overdot denotes differentiation with respect to proper time $\tau$.

In general, stationary and axisymmetric spacetimes, the metric tensor components are $t$- and $\phi$-independent thus admit at least two conserved quantities (due to stationarity and axisymmetry) throughout the geodesic evolution, namely the energy $E$ and $z$-component of the orbital angular momentum $L_z$ (see \cite{Destounis:2020kss,Destounis:2021mqv,Destounis:2021rko}).

The geometry \eqref{metric} is static and spherically-symmetric thus admits a third constants of motion, besides $E$ and $L_z$ which can be expressed from \eqref{metric} as
\begin{equation}\label{ELz}
	E/\mu=-g_{tt} \dot{t}, \qquad L_z/\mu=g_{\phi\phi}\dot{\phi},
\end{equation}
with $\mu$ the mass of the test-particle (secondary). The third constant corresponds to the square of the angular momentum $L^2=L_x^2+L_y^2+L_z^2$ (see Sec. \ref{integrable} and Appendix \ref{appA}). The $t$- and $\phi$-momenta can be expressed with respect to the conserved quantities and the non-zero metric tensor components. Together with the conservation of the rest mass $\mu$ of the secondary, (preservation of four-velocity) which leads to $g_{\lambda\nu}\dot{x}^\lambda\dot{x}^\nu=-1$, the geodesics of test particles possess four constants of motion. Specifically, the conservation of the rest mass gives the constraint equation for bound orbits
\begin{equation}\label{constraint}
	\dot{r}^2+\frac{g_{\theta\theta}}{g_{rr}}\dot{\theta}^2+V_\text{eff}=0,
\end{equation}
where the Newtonian-like effective potential has the form
\begin{equation}
	V_\text{eff}\equiv \frac{1}{g_{rr}}\left(1+\frac{g_{\phi\phi}E^2+g_{tt} L_z^2}{g_{tt}g_{\phi\phi}}\right).
\end{equation}
The curve defined when $V_\text{eff}=0$ is called the curve of zero velocity (CZV) since $\dot{r}=\dot{\theta}=0$ there. Utilizing the CZV and proper initial conditions leads to bound orbits that do not escape from the gravitational potential of the primary nor plunge into the BH.

Generally, bound geodesics can be fully characterized by three frequencies which are imprinted at the emitted gravitational radiation of EMRIs. These frequencies  are associated with the radial rate of transition between the periapsis and apoapsis of the orbit ($\omega_r$), longitudinal oscillations around the equatorial plane ($\omega_\theta$) and the frequency of revolution around the primary ($\omega_\phi$). The geodesics, then, evolve on two-dimensional tori characterized by the above frequencies. When the ratios $\omega_r/\omega_\theta$, $\omega_r/\omega_\phi$ or $\omega_\theta/\omega_\phi$, are irrational then the orbits are quasi-periodic and cover the entire phase space of the associated torus densely, meaning that they never return to their initial position. On the other hand, when one of these ratios form a rational number then the geodesic is periodic (or resonant) and returns to its initial position after a number of oscillations defined by the ratio. Such orbits are special in the sense that they are not phase-space filling and therefore, can directly affect the evolution of EMRIs \cite{Flanagan:2010cd,Flanagan:2012kg,Brink:2013nna,Ruangsri:2013hra,vandeMeent:2013sza,vandeMeent:2014raa,Brink:2015roa,Berry:2016bit,Speri:2021psr,Gupta:2022fbe,Apostolatos:2009vu,Lukes-Gerakopoulos:2010ipp,Zelenka:2019nyp,Lukes-Gerakopoulos2020,Mukherjee:2022dju,Destounis:2020kss,Destounis:2021mqv,Destounis:2021rko,Destounis:2023gpw}.

\subsection{Integrability}\label{integrable}

In general, the metric tensor field of geometry \eqref{metric} (and any other spherically-symmetric configuration) admits four Killing vector fields (KVFs); one timelike $\eta^{\alpha}$ which acts simply transitively, and three spacelike $\xi^{\alpha}_{(i)},\, i\in \{1,2,3\}$ which act multiple transitively in two-dimensional surfaces. Locally, the algebra of the three spacelike KVFs form an $\text{SO(3)}$ group, while they all commute with the timelike field. As discussed above, every KVF gives rise to a linear (in velocities) constant of motion for the geodesics equations \eqref{geodesic}. Thus, if
\begin{align}
	E=\eta^{\alpha}g_{\alpha\beta}\dot{x}^{\beta},\qquad
	L_{i}=\xi^{\alpha}_{(i)}g_{\alpha\beta}\dot{x}^{\beta},\,\, i\in \{1,2,3\},
\end{align}
then
\begin{equation}
	\dot{x}^{\alpha}\nabla_{\alpha}E=\dot{x}^{\alpha}\nabla_{\alpha}L_{i}=0,
\end{equation}
provided that Eq. \eqref{geodesic} is satisfied, i.e., four linear constants are expected. Given the fact that the initial degrees of freedom are four in Eq. \eqref{geodesic}, eight boundary or initial conditions are needed in order for the system to be integrated. Nevertheless, two out of the four linear constants,
i.e., the integrals $L_{2}$ and $L_{3}$, are functionally dependent, otherwise the system would be super-integrable and it could be solved algebraically. Therefore, only three out of the four integrals of motion can be implemented, suggesting that there is only one degree of freedom left that corresponds to the radial coordinate $r$. 

Alternatively, if one tries to search for a quadratic (in velocities) Killing tensor, e.g. $K_{\alpha\beta}$, then the corresponding equations $\nabla_{(\alpha}K_{\beta\gamma)}=0$ can easily be solved in the case under consideration. It turns out that
\begin{align*}
	K^{\alpha\beta}&=K_{1}\eta^{\alpha}\eta^{\beta}+K_{2}\big(\xi^{\alpha}_{(1)}\xi^{\beta}_{(1)}
	+\xi^{\alpha}_{(2)}\xi^{\beta}_{(2)}\big)
	+K_{3}\xi^{\alpha}_{(3)}\xi^{\beta}_{(3)}\\
	&+K_{4}\big(\eta^{\alpha}\xi^{\beta}_{(3)}
	+\xi^{\alpha}_{(3)}\eta^{\beta}\big)
	+K_{5}g^{\alpha\beta}.
\end{align*}
Since the constants $K_{j},\,j\in\{1,2,3,4,5\}$ are free parameters, one can set $K_{1}=K_{4}=K_{5}=0$ and $K_{2}=K_{3}=1$. Then the (would be) Carter constant equals to the magnitude of the angular momentum operator
\begin{equation}
	K_{\alpha\beta}\dot{x}^{\alpha}\dot{x}^{\beta}
	=L^{2}_{1}+L^{2}_{2}+L^{2}_{3}=L^{2}_{x}+L^{2}_{y}+L^{2}_{z}\equiv {L}^{2}.
\end{equation}
In that case, the needed integrals of motion, in order to perform a reduction in the geodesics, are $E,\, L_z$ and $L^{2}$, which can be solved algebraically in terms of the velocities $\dot{t},\dot{x},\dot{\phi}$ (where $x=\cos\theta$) and subsequently substituted (along with their first derivatives with respect the affine parameter) to the geodesics. Hence, a single, second-order ordinary differential equation regarding the radial coordinate $r$, will emerge (see Appendix \ref{appA}) that describes the radial evolution of geodesics on a fixed plane $\theta=\text{constant}$.

\begin{figure*}[t]
	\includegraphics[scale=0.31]{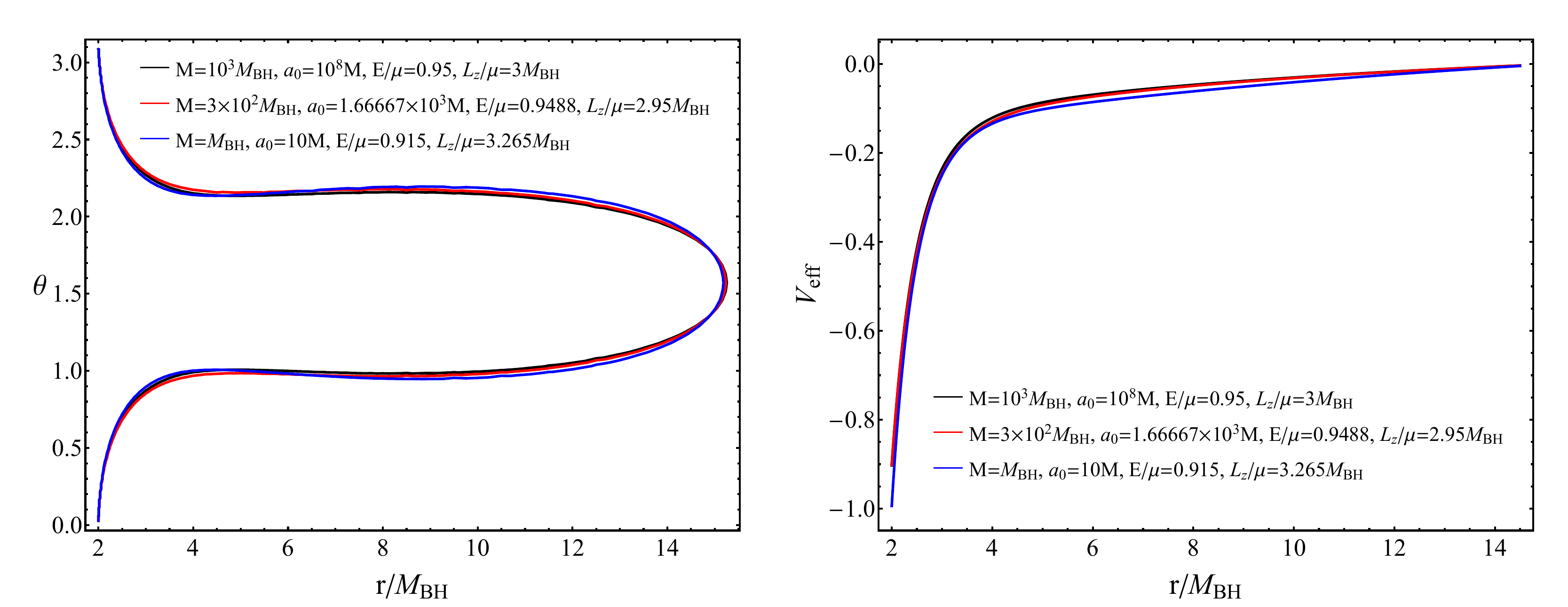}
	\caption{\emph{Left:} Curves of zero velocity for the geometry \eqref{metric} with various combinations of the parameters $E,\, L_z,\, M,\, a_0$. The compactness combinations vary in the range $M/a_0\in[10^{-8},10^{-1}]$. \emph{Right:} Same as the left figure with $\theta=\pi/2$. For both cases the secondary and primary masses are $\mu=2M_\odot$ and $M_\text{BH}=2\times10^6 M_\odot$, respectively.}\label{fig0}
\end{figure*}

\begin{figure*}[t]
	\includegraphics[scale=0.35]{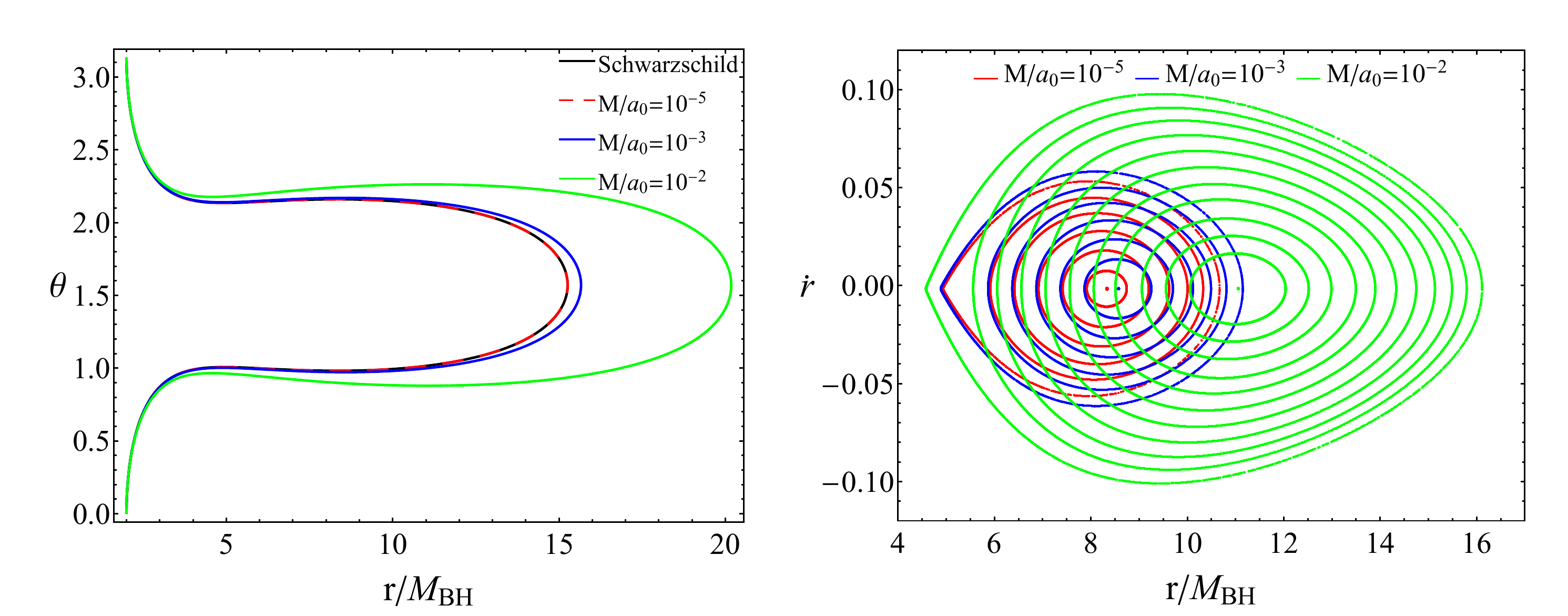}
	\caption{\emph{Left:} Curves of zero velocity for the geometry \eqref{metric} with $M=10^3 M_\text{BH}$ and varying compactness $M/a_0$. For comparison we also present the curve of zero velocity of Schwarzschild geodesics. \emph{Right:} Poincar\'e maps of bound orbits for the geometry \eqref{metric} with $M=10^3 M_\text{BH}$ and varying compactness $M/a_0$. For both cases the conserved energy and angular momentum of the test particle are chosen as $E/\mu=0.95$, $L_z/\mu=3M_\text{BH}$, respectively, where the secondary and primary masses are $\mu=2M_\odot$ and $M_\text{BH}=2\times10^6 M_\odot$, respectively.}\label{fig1}
\end{figure*}

\begin{figure*}
	\includegraphics[scale=0.35]{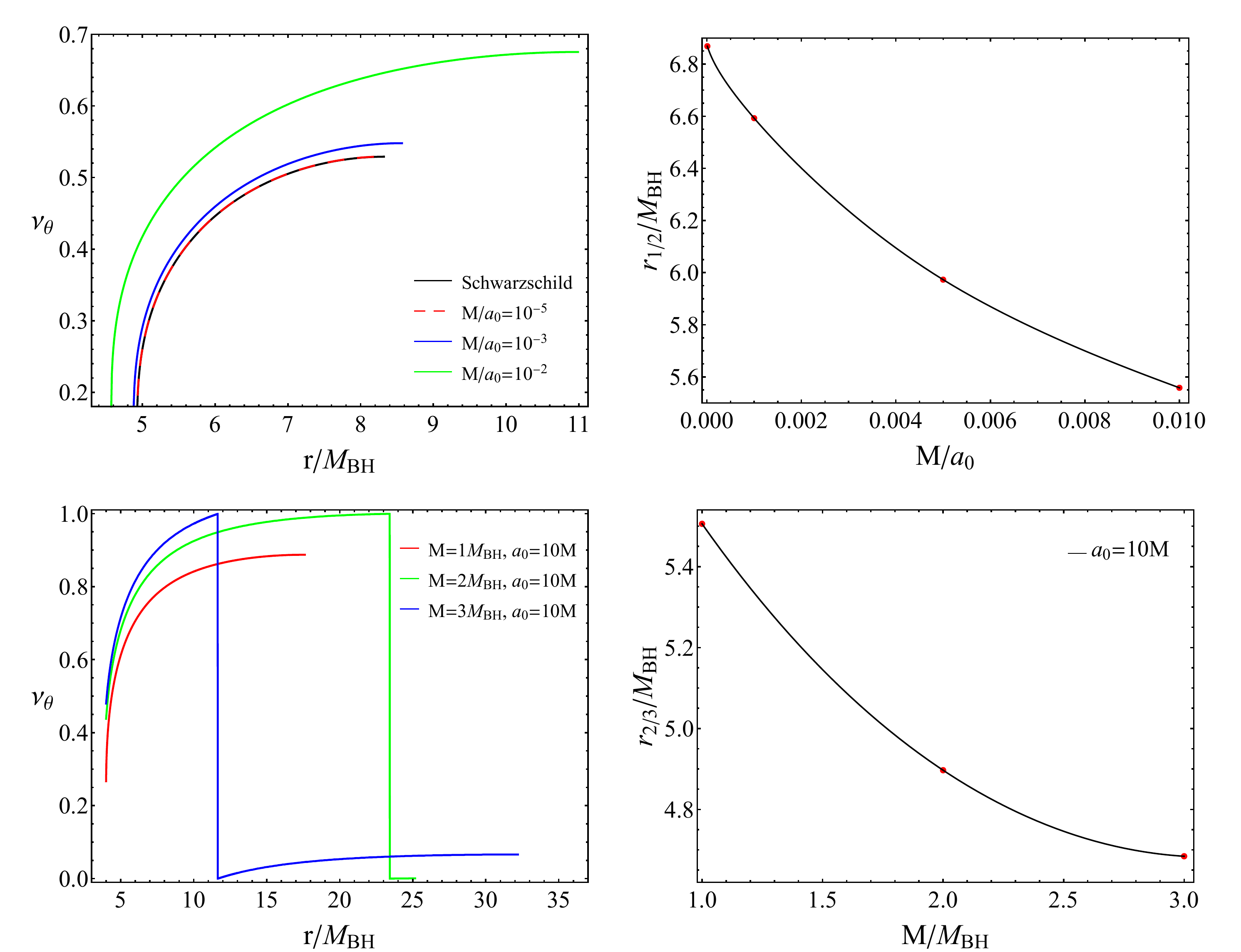}
	\caption{\emph{Top left:} Rotation curves of bound orbits for the geometry \eqref{metric} with $M=10^3 M_\text{BH}$ and varying compactness $M/a_0$. \emph{Top right:} Radial position of $1/2$-resonances $r=r_{1/2}M_\text{BH}$ with respect to the compactness $M/a_0$ extracted from the top left curves. \emph{Bottom left:} Rotation curves of bound orbits for the geometry \eqref{metric} with $a_0=10 M$ and varying halo mass $M$. \emph{Bottom right:} Radial position of $2/3$-resonances $r=r_{2/3}M_\text{BH}$ with respect to the halo mass $M$ extracted from the bottom left curves. For all cases, the conserved energy and angular momentum of the test particle are chosen as $E/\mu=0.95$, $L_z/\mu=3M_\text{BH}$, respectively, where the secondary and primary masses are $\mu=2M_\odot$ and $M_\text{BH}=2\times10^6 M_\odot$, respectively.}\label{fig2}
\end{figure*}

\begin{figure*}[t]
	\includegraphics[scale=0.42]{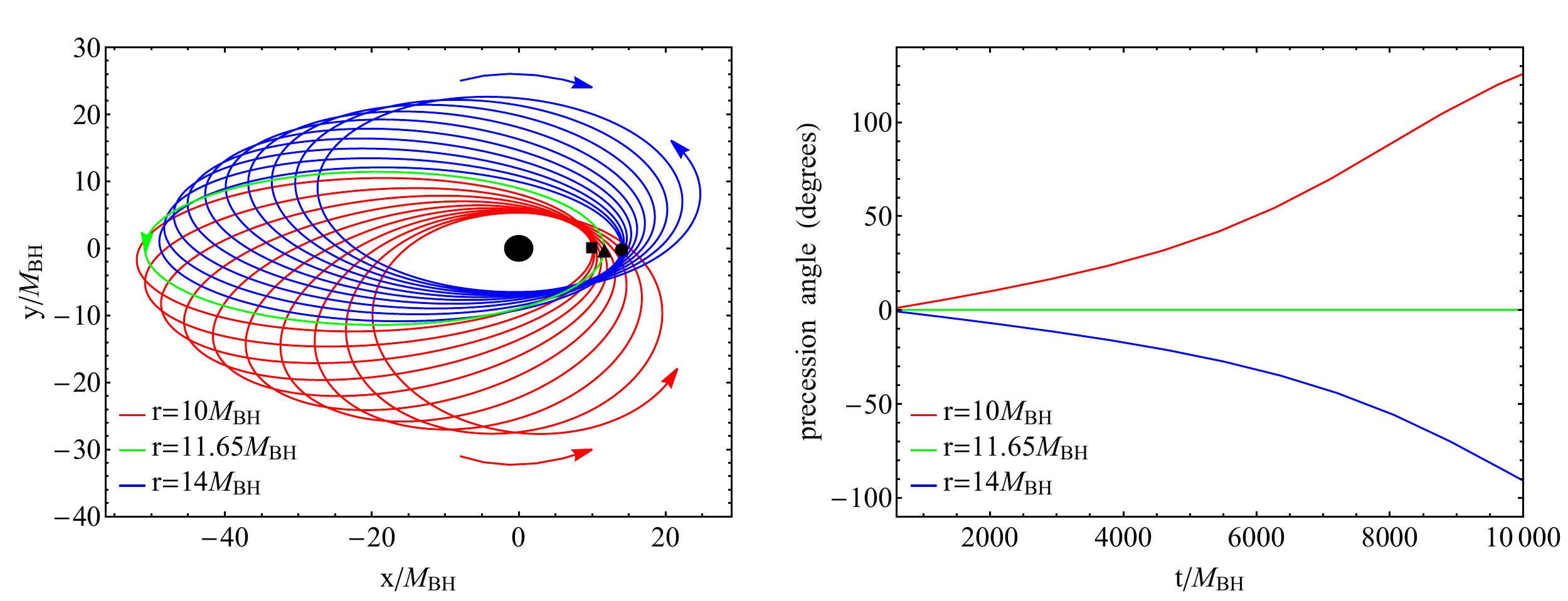}
	\caption{\emph{Left:} Geodesic evolution of a compact object with mass $\mu=2 M_\odot$ orbiting around a supermassive BH with mass $M_\text{BH}=2 \times 10^6 M_\odot$. The BH, shown at the origin, has event horizon radius $r=2 M_\text{BH}$ and resides in a compact environment with mass $M=3 M_\text{BH}$ and length scale $a_0=10M$, described by the geometry \eqref{metric}. Each orbit is evolved for ten revolutions and projected in Euclidean coordinates. Different colors designate orbits with distinct initial positions $r=r(0)$ as well as with $\dot{r}(0)=0,\,\theta(0)=\pi/2,\, E/\mu=0.95,\,L_z/\mu=3M_\text{BH}$, while $\dot{\theta}(0)$ is initialized through the constraint equation \eqref{constraint}. The geodesic with $r(0)=10M_\text{BH}$ (red) begins its trajectory from the black square in the $(x,y)$ plane. Equivalently, orbits with $r(0)=11.65M_\text{BH}$ (green) and $r(0)=14M_\text{BH}$ (blue) begin from the black triangle and circle, respectively. The red, green and blue arrowheads at the end of evolution designate the orbital direction which is anticlockwise for all cases. Finally, the red and blue arrows point towards the direction of precession. \emph{Right:} Evolution of precession angles for the three aforementioned orbital cases with respect to time for $\sim 50$ cycles.}\label{fig3}
\end{figure*}

\begin{figure}
	\includegraphics[scale=0.38]{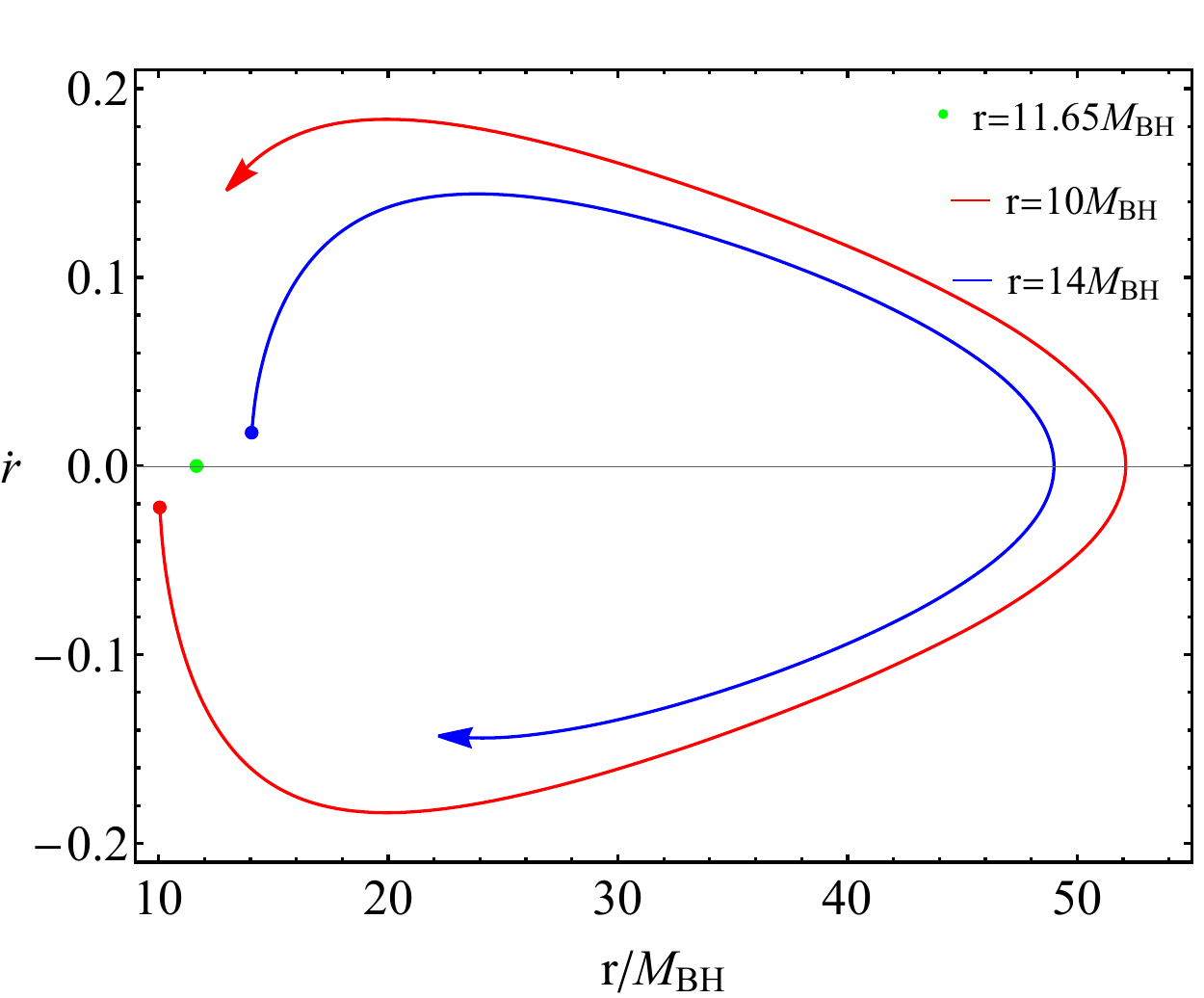}
	\caption{Equatorial surface of section of geodesics with initial conditions $r(0)=10M_\text{BH}$ (red), $r(0)=11.65M_\text{BH}$ (green), $r(0)=14M_\text{BH}$ (blue), respectively, and $\dot{r}(0)=0,~\theta(0)=\pi/2,~E/\mu=0.95,~L_z/\mu=3M_\text{BH}$, while $\dot{\theta}(0)$ is initialized through the constraint equation \eqref{constraint}. The secondary (with $\mu=2 M_\odot$) orbits around a primary (with $M_\text{BH}=2 \times 10^6 M_\odot$) that is described by the geometry \eqref{metric} with $M=3M_\text{BH}$ and $a_0=10M$. Here, only the first $50$ intersections through the equatorial plane are shown to emphasize the change in intersection direction.}\label{fig4}
\end{figure}

\subsection{Poincar\'e surface of section and rotation number}

To comprehend the structure of bound orbits in phase space around the geometry \eqref{metric} we can employ various tools in order to gain further intuition regarding interesting orbital phenomena. A typical example is the Poincar\'e map which is constructed by successive intersections of orbits, with varying initial conditions, on a surface of section (e.g. the equatorial plane) with positive (or negative) direction of intersection. The structure of the Poincar\'e map can instantly reveal the existence of chaos, such as disorganized intersections which reveal a fully chaotic/ergodic orbital evolution or the formation of resonant/Birkhoff islands that encapsulate periodic-orbit stable points \cite{Contopoulos_book}. To further elaborate on the libration-like frequency evolution, one can utilize the rotation number of each geodesic. This is accomplished by tracking the angle $\vartheta$ between two successive intersections on the Poincar\'e map relative to the fixed central point of the map (clockwise or anticlockwise) which corresponds to a circular, but otherwise not necessarily equatorial, orbit that intersects the surface of section exactly at the same point. The rotation number is then defined as \cite{Contopoulos_book}
\begin{equation}
	\nu_\vartheta=\frac{1}{2 \pi N}\sum_{i=1}^{N}\vartheta_i.
\end{equation}
When the number of angles measured $N$ tends to infinity, the above sequence converges to the rotation number $\nu_\vartheta=\omega_r/\omega_\theta$. Integrable systems (such as the one we study here) exhibit monotonous changes in consecutive rotation numbers. Rotation curves formed by successive rotation numbers are a rather helpful tool to spot where resonances lie and if there are any imprints of chaos in dynamical systems \cite{Contopoulos_book,Apostolatos:2009vu,Lukes-Gerakopoulos:2010ipp,Lukes-Gerakopoulos:2012qpc,Lukes-Gerakopoulos:2013gwa,Contopoulos:2011dz,Lukes-Gerakopoulos:2014dpa,Destounis:2020kss,Destounis:2021mqv,Destounis:2021rko,Destounis:2023gpw}.

\section{Environmental effects on geodesic evolution}

In this section we perform a qualitative comparison between geodesics around vacuum Schwarzschild BHs and those evolving in the geometry \eqref{metric}. By solving the coupled radial and polar second-order ordinary differential equations (without making any assumptions of integrability), together with the first order decoupled equations for $\dot{t}$ and $\dot{\phi}$ from Eqs. \eqref{ELz}, we obtain bound orbits that reside inside the CZV and never plunge nor escape to infinity\footnote{It is important to note that since the spacetime under consideration is integrable, one can decouple the equations for the evolution of $r$ and $\theta$ to become separate first-order differential equations. Nevertheless, symmetry assumptions only simplify the equations meaning that one can integrate them faster but the resulting orbits are identical in both cases.}. This is made possible by the use of appropriate initial velocity components for $\dot{r},\,\dot{\theta}$. All orbits we obtained lie on a fixed plane, as expected from spherical symmetry and the discussion in Sec. \ref{integrable}. To check the precision of our evolution we evolve the constraint equation \eqref{constraint} for $10^4-10^5$ revolutions and find that it is satisfied within one part in $10^{10}-10^{12}$ depending on the compactness of the halo.

Before embarking in a parametric space analysis of geodesics we note that there is a variety of parameter sets $(E/\mu,L_z/\mu,M,a_0)$ that can give rise to almost identical CZVs and $V_\text{eff}$ at the equatorial plane. In Fig. \ref{fig0} we demonstrate the aforementioned statement for three different cases of $(E/\mu,L_z/\mu,M,a_0)$ sets that give rise to similar (if not the same) potentials and CZVs. Note that we have spanned the compactness of the halo in a rather large range, i.e. $M/a_0\in\left[10^{-8},10^{-1}\right]$, and could still find appropriate choices of $E$ and $L_z$ that lead to similar orbital potentials. Since the parameter space $(E/\mu,L_z/\mu,M,a_0)$ presents such degeneracies, we will fix $E/\mu=0.95$ and $L_z/\mu=3M_\text{BH}$ for the rest of the paper. The particular choice of $E$ and $L_z$ give rise to geodesics with small eccentricity, generally. Due to the fact that most of our initial conditions lie in the strong field regime and close to the primary, where the emission of GWs have significantly circularized the orbits, this particular set of initial parameters for the secondary are astrophysically relevant.

\subsection{Poincar\'e maps, rotation curves and resonances}

The predominant effect introduced by the halo is an overall redshift on the fundamental structure of the geometry, such as the light-ring position, as well as a redshift on the light-ring angular frequency, the instability timescale of null geodesics and the characteristic vibrational frequencies (quasinormal modes) of spacetime under scalar and axial gravitational perturbations \cite{Cardoso:2021wlq}.

The geodesic analysis reveals, at first glance, a volume enlargement of the CZVs, with respect to that of Schwarzschild, as the compactness of the halo increases (see Fig. \ref{fig1}). Such behavior translates to orbits that can span on a larger orbital frequency range which is imprinted in the Poincar\'e surface of section shown in Fig. \ref{fig1}. Galactic-scale halos with compactness of order $M/a_0\lesssim 10^{-5}$ do not seem to affect significantly the available orbital phase space of bound orbits at first glance. As we will see later though, even compactnesses of astrophysical relevance can affect the geodesic evolution and the emitted GWs substantially, as has already been shown in \cite{Cardoso:2021wlq}, though only for circular equatorial EMRIs. Nevertheless, upgrading the halo into a compact BH hair-like environment leads to a significant change on the bound orbit phase space. Fig. \ref{fig2} portrays the orbital effect of the halo for small and intermediate compactness (top panel), as well as dense BH environments (bottom panel). 

The rotation curves (designating the frequency ratio of radial and polar orbital oscillations) for small and intermediate compactness have a trivial monotonic structure which agrees with the integrability property of geodesics in spherical symmetry. The main effect presented on the top panel of Fig. \ref{fig2} is a redshift of the orbital frequency ratio into smaller radii with increasing compactness, with respect to that of vacuum Schwarzschild geodesics, as well as the sustainability of orbits with higher rotation numbers. A qualitative picture is presented for a particular resonant frequency, namely the $\omega_r/\omega_\theta=1/2$ periodic orbit (top right panel in Fig. \ref{fig2}) and the $\omega_r/\omega_\theta=2/3$ periodic orbit (bottom right panel in Fig. \ref{fig2}), which occurrence experiences an advance towards the primary as the halo becomes more dense. One may characterize such events typical due to the growing presence of gravitating dark matter around the primary. This is indeed the case; the increment of compactness leads to an antagonism between the gravitational pull of the primary and the dark matter influence on test particles which allows for bound orbits closer to plunge and further regions before escaping to infinity, as well as with higher orbital frequency ratios. If there would be a case where geodesics around a non-vacuum primary mimic the rotation curve of orbits around a vacuum Kerr BH, then such degeneracy can easily be broken directly from the properties of resonances in spherical symmetry, which do not affect the fluxes, in contrast to Kerr resonances which have been shown to affect significantly the resulting waveforms and the fluxes of energy, $z$-component of angular momentum and Carter constant \cite{Flanagan:2010cd,Flanagan:2012kg,Berry:2016bit}, as well as parameter estimation \cite{Speri:2021psr,Gupta:2022fbe}.

\subsection{Apsidal precession drift}\label{orb_prec}

Intriguingly, when introducing a more compact environment surrounding the primary object the dynamics display an interesting phenomenon, related to the apsidal precession of the orbit. In Fig. \ref{fig2} (bottom left panel) we show rotation curves of a BH surrounded by a compact environment with fixed $M/a_0=10^{-1}$ and increasing halo mass $M$. Beyond a certain halo mass the rotation curve can reach unity at a critical radius after which the rotation number drops from unity to zero, and then slightly increases to non-zero values. Such event is not a numerical artifact\footnote{We have performed intense convergence tests with increasing number of intersections and initial conditions around the critical radius and always retrieve the same discontinuity up to numerical precision.} but rather a physical phenomenon related to the antagonism and eventual counterbalance between general-relativistic effects and the gravitational field of dark matter. As we will see below, such physical antagonism will lead to a change in the precession drift direction which leads to these critical radii in rotation curves. Therefore, what occurs in Fig. \ref{fig2} depends on the definition of the rotation number, which is directly linked to the assumption made regarding the direction one measures angles between successive intersections.

Fig. \ref{fig3} (left panel) portrays three distinct EMRI orbits with different initial radial positions in the $x-y$ plane under the identification of Boyer-Lindquist coordinates with spherical coordinates as seen from an observer at infinity. The orbit in red initiates its trajectory in a region before the critical radius. In this case the precession drift is positive (prograde precession), i.e. the orbit's axis rotates in the same direction as the orbital motion. On the other hand, the orbit in blue, with initial radial position beyond the critical point, has negative precession (retrograde precession) since the apsidal axis rotates in the opposite direction as the trajectory of the test particle. Right at the critical radius where the drop on the rotation curves occur lies an elliptic orbit with a critical initial condition $r(0)$ that does not exhibit precession in any direction (shown in green in Fig. \ref{fig3}). On the right panel of Fig. \ref{fig3} we present the precession angle evolution for the three aforementioned geodesics which clearly demonstrates that the retrograde and prograde trajectories have opposite precession rates while the elliptic orbit's rate remains constant, i.e. exhibits no precession drift.

Therefore, even though at first glance the rotation curves present severe discontinuities, they can be explained by physical phenomena and happen smoothly as the initial secondary's precession rate slowly decreases, becomes null and eventually changes sign, with the decrease of the initial radial position of the geodesic. Discontinuities in rotation curves can therefore be associated with a change in the precession drift's direction from prograde to retrograde and eventually trace back to the particular conventions made in the definition of the rotation number.

Indeed, by inspecting the consecutive intersections through a surface of section of the orbits presented in Fig. \ref{fig3}, we find that the intersections of the retrograde- and prograde-precessing geodesics have opposite direction, while the non-precessing orbit intersects the surface of section at exactly one single point (see Fig. \ref{fig4}). Since the rotation number depends on the direction which one measures the angles between consecutive crossings (clockwise or counterclockwise), the rotation curve is bound to drop to zero when the critical radius is met, where the angle between subsequent intersections is zero. Physically, the trajectories close to the primary are dominated by general-relativistic effects and the precession drift is positive (similar to what occurs in the precession of Mercury and the S-stars around Sgr A*), while beyond the critical point, where the gravitational field of dark matter is dominant, the dynamics of test particles and precession rates become negative.

Finally, we note that, to our knowledge, the aforementioned phenomenon only occurs when dark matter \cite{Igata:2022nkt,Igata:2022rcm}, other novel fundamental fields that change the theory of gravity \cite{Chatzifotis:2022ene}, thick accretion disks \cite{Murray:2022}, as well as exotic compact objects, such as wormholes \cite{Potashov:2020zmg}, are considered. Geodesics around Schwarzschild BHs always precess prograde; a phenomenon which has also been observed experimentally on the S-stars that orbit around the supermassive BH in the center of the Milky Way \cite{GRAVITY:2020gka}, which according to contemporary estimations of its spin is rather slowly-rotating, thus can be modeled as a Schwarzschild BH \cite{Melia:2001fp,Fragione:2020khu}.

\section{Gravitational radiation}

Even though we operate at the geodesic level, and do not take into account radiative backreaction of the secondary to the geometry of the primary, it is still interesting to qualitatively compare the approximate GW emission of a particle in vacuum Schwarzschild and a particle orbiting a Schwarzschild BH immersed in a dark matter halo where dynamical friction actively takes place. For this task, we shall take advantage of the quadrupole approximation described below.

\begin{figure*}[t]
	\includegraphics[scale=0.34]{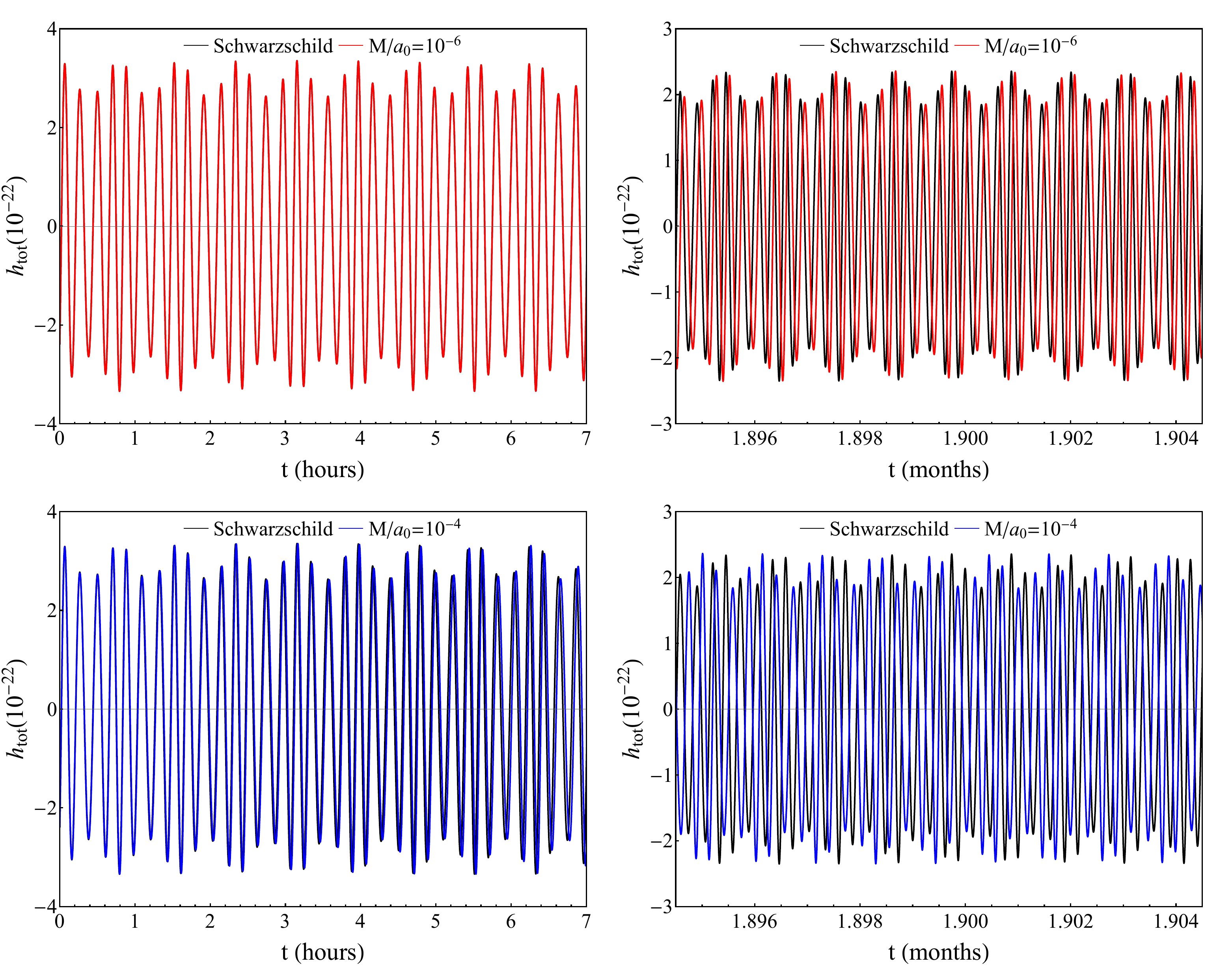}
	\caption{\emph{Top panel:} GWs detected by LISA at the beginning (left subfigure, first 7 hours) and at the end (right subfigure, after $~2$ months) of an EMRI composed of a vacuum Schwarzschild primary (black curves) and primary surround by a halo with $M/a_0=10^{-6}$ (red curves), where $\mu=2M_\odot$, $M_\text{BH}=2\times 10^6 M_\odot$. The halo mass is chosen as $M=10^3M_\text{BH}$. The secondary is initialized for all cases with $r(0)=7.5M_\text{BH}$, $\dot{r}(0)=0$, $\theta(0)=\pi/2$, $E/\mu=0.95$, $L_z/\mu=3M_\text{BH}$, while $\dot{\theta}(0)$ is found from the constraint equation \eqref{constraint}. The waveforms result from the orbital evolution of the aforementioned geodesics for $t=5\times 10^5 M_\text{BH}\sim 2$ months (or $4\times10^3$ orbital revolutions) and take only into account the $\ell=2$ contribution to gravitational radiation. The corresponding radiation is observed by LISA from luminosity distance $d=100 \text{Mpc}$. \emph{Bottom panel:} Same as the top panel for an EMRI composed of a vacuum Schwarzschild primary (black curves) and a primary surround by a halo with $M/a_0=10^{-4}$ (blue curves).}\label{fig4.1}
\end{figure*}

\begin{figure}[t]
	\includegraphics[scale=0.355]{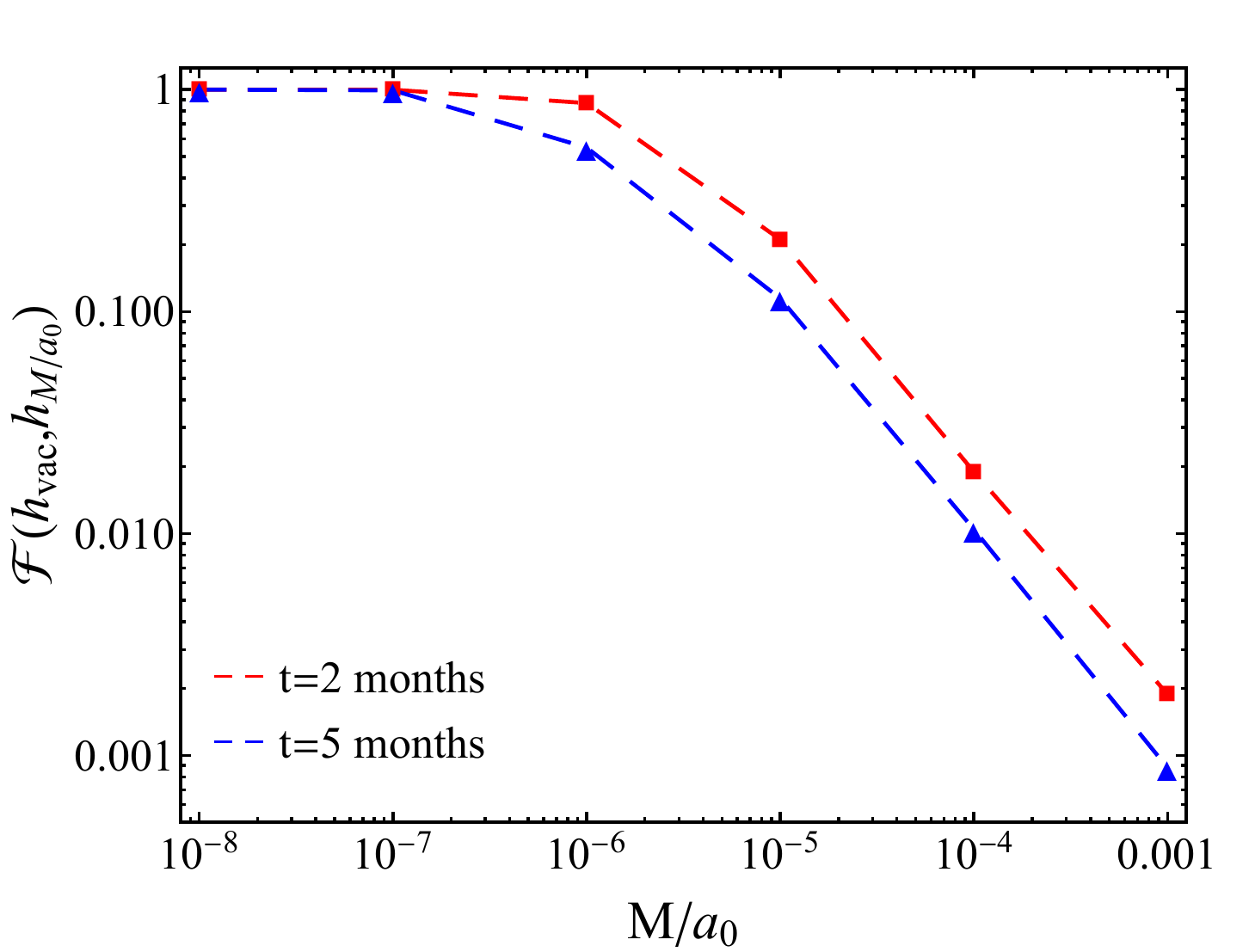}
	\caption{Faithfulness $\mathcal{F}(h_1,h_2)$ (in logarithmic scale) between GWs from a vacuum Schwarzschild EMRI $h_\text{vac}$ and non-vacuum EMRI with varying halo compactness $h_{M/a_0}$, where the halo mass is set to $M=10^3M_\text{BH}$ and the primary BH's mass $M_\text{BH}=2\times 10^6 M_\odot$. The secondary has mass $\mu=2M_\odot$ and is initialized for all cases with $r(0)=7.5M_\text{BH}$, $\dot{r}(0)=0$, $\theta(0)=\pi/2$, $E/\mu=0.95$, $L_z/\mu=3M_\text{BH}$, while $\dot{\theta}(0)$ is found from the constraint equation \eqref{constraint}. The faithfulness between the waveforms is calculated for the twn (red curve) and five (blue curve) months of observation.}
	\label{fig4.2}
\end{figure}

\begin{figure*}[t]
	\includegraphics[scale=0.38]{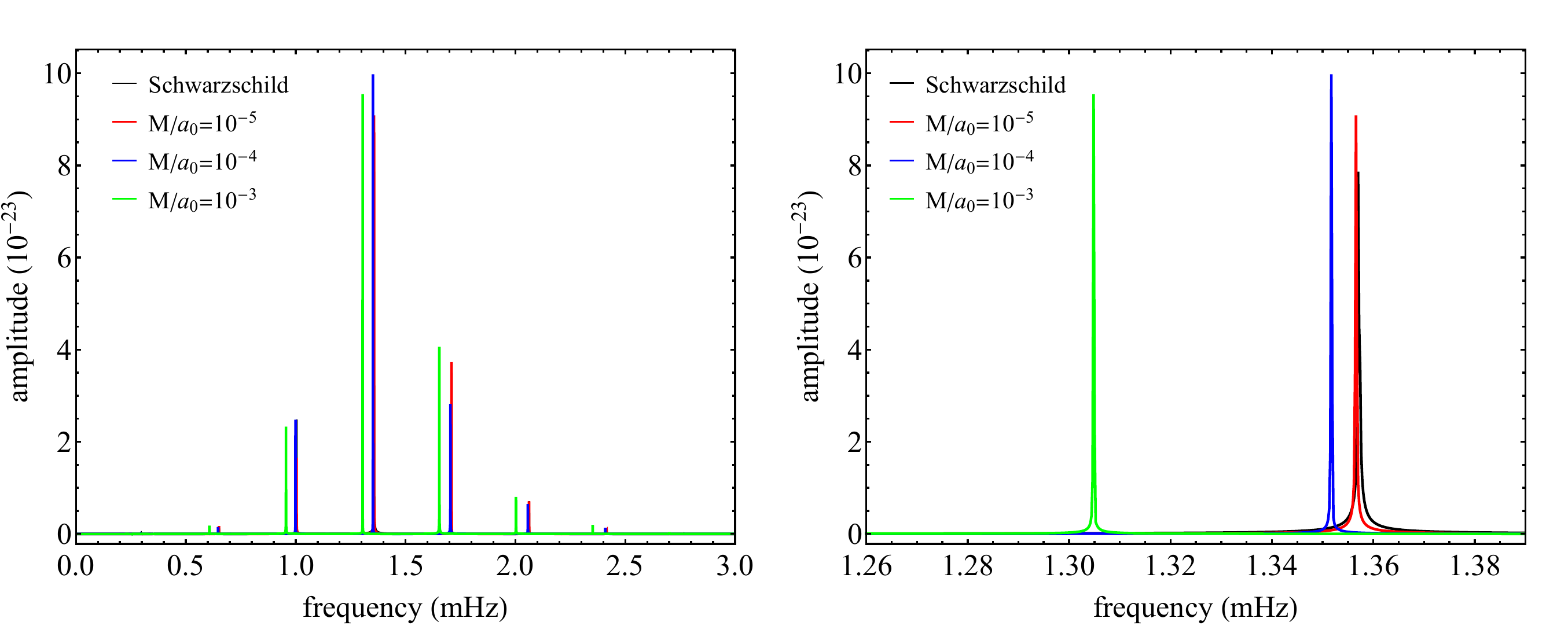}
	\caption{\emph{Left:} GW frequencies (in mHz) extracted from the waveforms produced by EMRIs, and observed by LISA, with $\mu=2M_\odot$, $M_\text{BH}=2\times 10^6 M_\odot$ and different compactness $M/a_0$ where the halo mass is $M=10^3M_\text{BH}$. The secondary is initialized for all cases with $r(0)=7.5M_\text{BH}$, $\dot{r}(0)=0$, $\theta(0)=\pi/2$, $E/\mu=0.95$, $L_z/\mu=3M_\text{BH}$, while $\dot{\theta}(0)$ is found from the constraint equation \eqref{constraint}. The waveforms result from the orbital evolution of the aforementioned geodesics for $t=2\times 10^6 M_\text{BH}\sim7.5$ months (or $1.5\times 10^4$ orbital revolutions) and take only into account the $\ell=2$ contribution to gravitational radiation. The corresponding radiation is observed by a space-based detector from luminosity distance $d=100 \text{Mpc}$. \emph{Right:} Zoom into a particular region of the left figure to discern the redshift effect that compactness introduces to the waveforms.}\label{fig5}
\end{figure*}

\begin{figure*}[t]
	\includegraphics[scale=0.38]{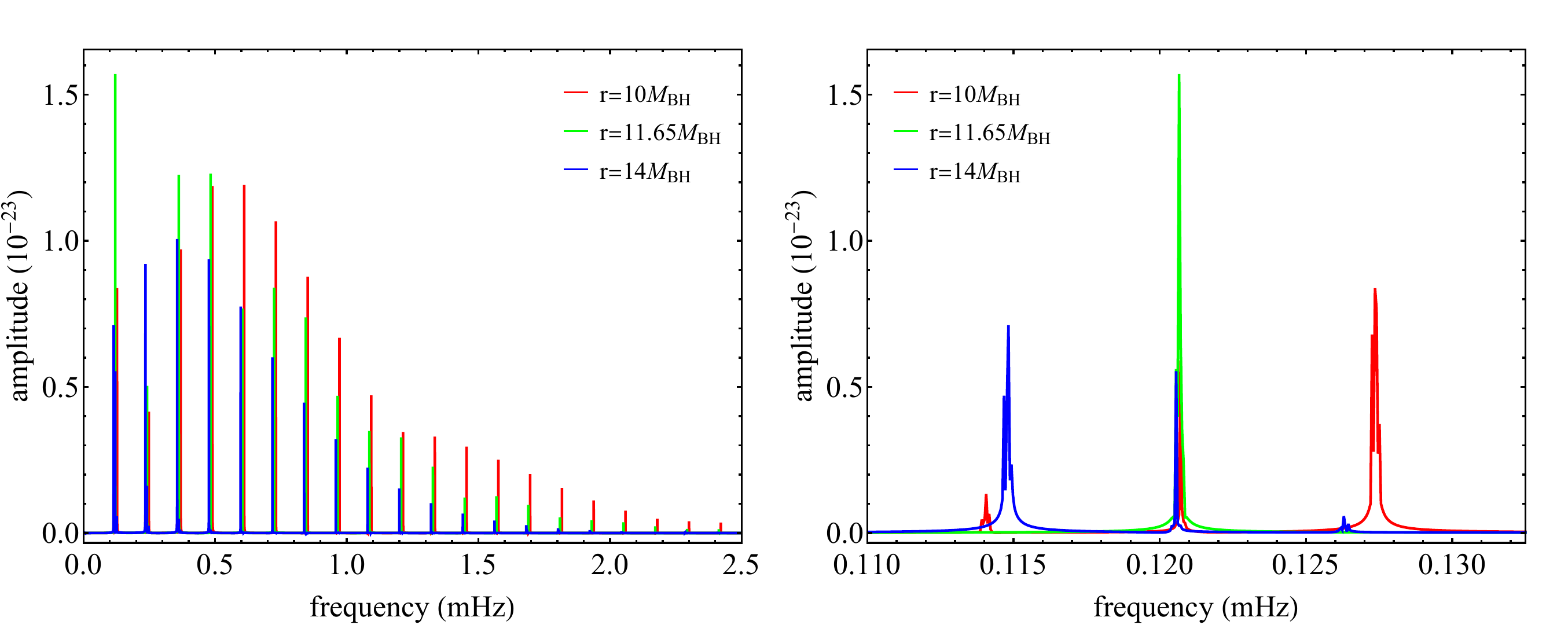}
	\caption{\emph{Left:} GW frequencies (in mHz) extracted from the waveforms produced by EMRIs, and detected by LISA, with $\mu=2M_\odot$, $M_\text{BH}=2\times 10^6 M_\odot$ and compactness $M/a_0=0.1$ where $M=3M_\text{BH}$ and $a_0=10M$. The secondary is initialized with $r(0)=10M_\text{BH}$ (red), $r(0)=11.65M_\text{BH}$ (green) and  $r(0)=14M_\text{BH}$ (blue) where $\dot{r}(0)=0$, $\theta(0)=\pi/2$, $E/\mu=0.95$, $L_z/\mu=3M_\text{BH}$, while $\dot{\theta}(0)$ is found from the constraint equation \eqref{constraint}. The waveforms result from the orbital evolution of the aforementioned geodesics for $t=2\times 10^6 M_\text{BH}\sim7.5$ months (or $1.5\times 10^4$ orbital revolutions) and take only into account the $\ell=2$ contribution to gravitational radiation. The corresponding radiation is observed by a space-based detector from luminosity distance $d=100 \text{Mpc}$. \emph{Right:} Zoom into a particular region of the left figure to discern the effect of precession drift reversal that compact environments introduce to the waveform.}
	\label{fig6}
\end{figure*}

\begin{figure*}[t]
	\includegraphics[scale=0.425]{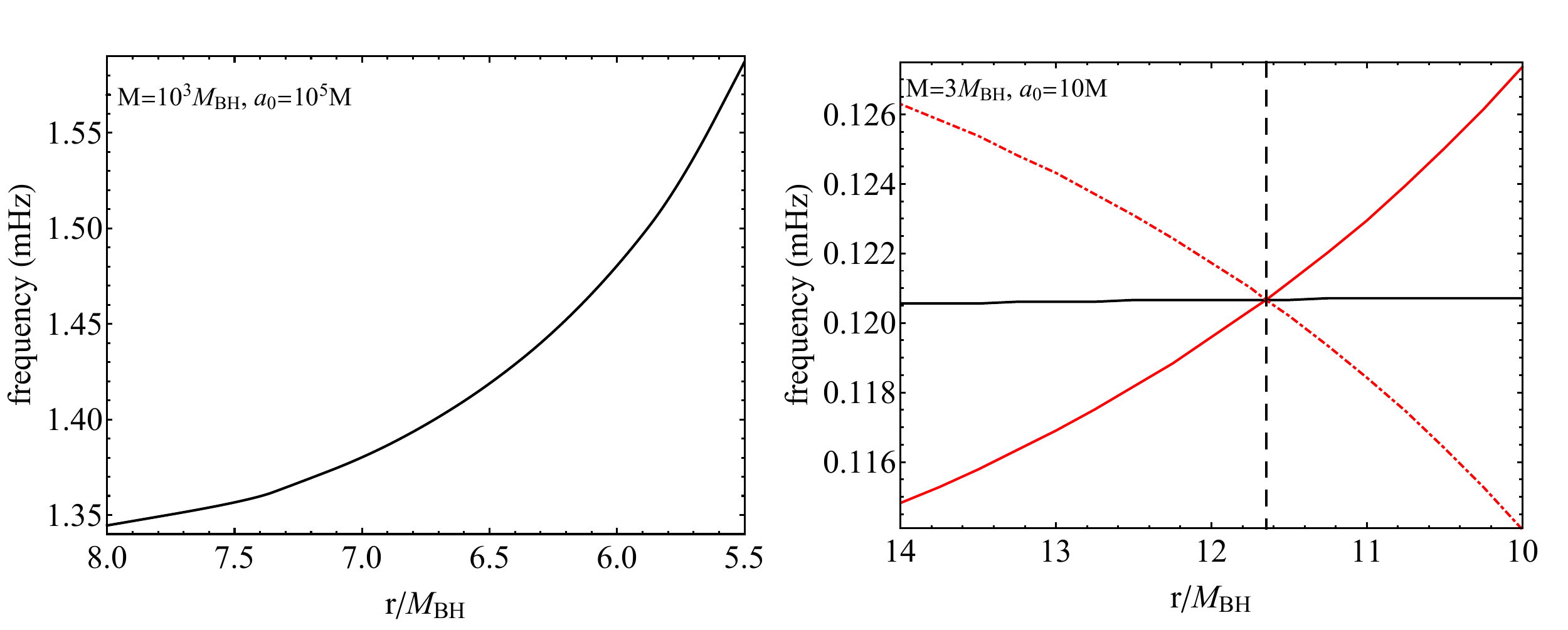}
	\caption{\emph{Left:} GW frequency (in mHz) with respect to the orbital radius $r(0)/M_\text{BH}$ extracted from waveforms produced by extreme-mass-ratio binaries with $\mu=2M_\odot$, $M_\text{BH}=2\times 10^6 M_\odot$ and compactness $M/a_0=10^{-5}$ where $M=10^3M_\text{BH}$ and $a_0=10^5M$. The secondary is initialized with $\dot{r}(0)=0$, $\theta(0)=\pi/2$, $E/\mu=0.95$, $L_z/\mu=3M_\text{BH}$, while $\dot{\theta}(0)$ is found from the constraint equation \eqref{constraint}. The waveforms result from the orbital evolution of geodesics for $t=2\times 10^6 M_\text{BH}\sim7.5$ months ($1.5\times10^4$ orbital revolutions) and take only into account the $\ell=2$ contribution to gravitational radiation. The corresponding radiation is observed by a space-based detector from luminosity distance $d=100 \text{Mpc}$. \emph{Right:} Same as left with $M/a_0=10^{-1}$ where $M=3M_\text{BH}$ and $a_0=10M$. In the observed frequency range, there exist three distinct peaks that evolve with $r/M_\text{BH}$. The red (and dashed-dotted red) curves show the dependence of the leftmost and outmost subpeaks to the initial condition $r(0)$ while the black line corresponds to the evolution of the central subpeak. The vertical dashed line designates the radius where the three subpeaks merge and the orbit is non-precessing.}\label{fig7}
\end{figure*}

\subsection{Quadrupole approximation}

The quadrupole formula takes advantage of the fact that the quadrupole emission of gravitational radiation is the dominant one, thus the radiative component of the metric perturbation introduced by the test particle at luminosity distance $d$ from the source $\boldmath{T}$ can be read at the transverse and traceless gauge as
\begin{equation}\label{metpert}
	h^\text{TT}_{ij}=\frac{2}{d}\frac{d^2 Q_{ij}}{dt^2},
\end{equation}
where $Q_{ij}$ is the symmetric and trace-free (STF) quadrupole tensor
\begin{equation}
	Q^{ij}=\left[\int x^i x^j T^{tt}(t,x^i) \,d^3 x\right]^\text{STF},
\end{equation}
with $t$ being the coordinate time measured at very large distances from the source. The source term of the point particle is then
\begin{equation}\label{Ttt}
	T^{tt}(t,x^i)=\mu \delta^{(3)}\left[x^i-Z^i(t)\right],
\end{equation}
where $Z(t)=(x(t),y(t),z(t))$ with 
\begin{align}
    x(t)&=r(t)\sin\theta(t)\cos\phi(t),\\
    y(t)&=r(t)\sin\theta(t)\sin\phi(t),\\
    z(t)&=r(t)\cos\theta(t),
\end{align} 
the trajectory components with respect to flat spherical coordinates, under the assumption that our space-borne detector is positioned at infinity. Even though this does not practically occur\footnote{Here, we identify the Schwarzschild coordinates $(r,\theta,\phi)$ of the secondary's trajectory with flat-space coordinates, known as the ``particle-on-a-string'' approximation.}, since we assume a finite luminosity distance $d$ from the source, such prescription, though not strictly valid, has been found to work well when generating EMRI waveforms in GR \cite{Babak:2006uv}.

An incoming GW onto the detector can be projected on its two polarizations, $+$ and $\times$, with the introduction of two unit vectors, namely $\boldsymbol{p}$ and $\boldsymbol{q}$, which are defined in terms of a third unit vector $\boldsymbol{n}$ that points from the source to the direction of the detector. The triplet of unit vectors $\boldsymbol{p},\,\boldsymbol{q},\,\boldsymbol{n}$ is chosen so that they form an orthonormal basis. The polarization tensor components are then
\begin{equation}
	\epsilon_+^{ij}=p^i p^j-q^i q^j,\,\,\,\,\,\,\epsilon_\times^{ij}=p^i q^j+p^j q^i,
\end{equation}
and allow us to write the metric perturbation as
\begin{equation}
	h^{ij}(t)=\epsilon_+^{ij}h_+(t)+\epsilon_\times^{ij}h_\times(t),
\end{equation}
with
\begin{equation}
	h_+(t)=\frac{1}{2}\epsilon_+^{ij}h_{ij}(t),\,\,\,\,\,\,\,h_\times(t)=\frac{1}{2}\epsilon_\times^{ij}h_{ij}(t).
\end{equation}
To express the GW components in terms of the position, $Z^i(t)$, velocity, $v^i(t)=dZ^i/dt$, and acceleration vectors $a^i(t)=d^2Z^i/dt^2$, we use Eqs. \eqref{metpert} and \eqref{Ttt} to obtain \cite{Canizares:2012is}
\begin{equation}
	\label{GW_formula}
	h_{+,\times}(t)=\frac{2\mu}{d}\epsilon^{+,\times}_{ij}\left[a^i(t)Z^j(t)+v^i(t)v^j(t)\right].
\end{equation}

LISA's response to an incident GW depends on the antennae patterns $F^{+,\times}_{I,II}$ of the detector (see Refs. \cite{Cutler:1997ta,Barack:2003fp,Destounis:2020kss} for their intricate functional forms), thus the total waveform detected by LISA is
\begin{equation}\label{total_GW}
	h_\alpha(t)=\frac{\sqrt{3}}{2}\left[F^+_\alpha(t) h_+(t)+F^\times_\alpha(t) h_\times(t)\right],
\end{equation}
where $\alpha=\{I,II\}$ is the channel index of the detector's antennae. We will simplify our analysis by assuming a detector that lies at a luminosity distance $d$ with fixed orientation $\boldsymbol{n}=(0,0,1)$ with respect to the source and that the primary's polar and azimuthal angles are fixed at the equatorial plane due to spherical symmetry (this choice simplifies a lot the response patterns of the antennae). 
	
A typical data stream observed by a detector contains both the signal of the source and some noise, but in our case we will assume that the noise is stationary and Gaussian with zero mean. Furthermore, we assume that the two data streams are uncorrelated and the noise power spectral density of LISA $S_n(f)$ (that includes instrumental, galactic and extra-galactic confusion noise \cite{Cutler:1997ta,Barack:2003fp}) is the same at both channels. This allows for a single-channel approximation. For more details we refer the reader to Refs. \cite{Cutler:1997ta,Barack:2003fp,Canizares:2012is,Destounis:2020kss}.

\subsection{Fourier analysis and waveform comparison}

Equation \eqref{total_GW} provides a decent approximate of the GWs emitted by a point-like particle orbiting around a supermassive primary and detected by LISA. Even though waveforms are obtained in the time domain, there exist a handful of data analysis schemes to maximize the phenomenological yield from GW observations.

The most significant tools in signal processing is usually connected to the Fourier transform of the signal from the time to the frequency domain. In what follows, we denote time domain waveforms as $h(t)$ and frequency domain ones, after being Fourier transformed, as $\tilde{h}(f)$, where $f$ is the frequency. A Fourier-transformed signal is by its nature represented with imaginary numbers, therefore whenever needed, we will take its absolute value in order to present figures of the Fourier peaks, and thus the resulting spectrum of GW signals.

When we want to answer questions regarding signal characterization, we can employ further statistical tests, in order to better understand the evolution of phase \cite{Owen:1995tm,Moore:2014lga}. The maximized overlap \cite{Owen:1995tm,McWilliams:2010eq,Hu:2022rjq}, or faithfulness, is a useful statistic for detailed waveform comparisons, since it is very sensitive to small differences in phase between signals. The faithfulness of two GWs is defined as the maximized noise-weighted overlap \cite{Owen:1995tm,Moore:2014lga}
\begin{equation}\label{faith}
       \mathcal{F}(h_1,h_2)=\max_{\{t_c,\Phi_c\}}\frac{\left<h_1|h_2\right>}{\sqrt{\left<h_1|h_1\right>\left<h_2|h_2\right>}},
\end{equation}
with $t_c$ and $\Phi_c$ being time and phase offsets. The the inner product $\left<h_1|h_2\right>$ is defined as
\begin{align}\nonumber
	\left<h_1|h_2\right>&=2\int_{f_\text{min}}^{f_\text{max}}\frac{\tilde{h}_1^*(f)\tilde{h}_2(f)+\tilde{h}_1(f)\tilde{h}_2^*(f)}{S_n(f)}df,\\\label{inner product}
	&=4\,\text{Re}\left[\int_{f_\text{min}}^{f_\text{max}}\frac{\tilde{h}_1^*(f)\tilde{h}_2(f)}{S_n(f)}df\right],
\end{align}
where the superscript $*$ designates complex conjugation, and the Fourier transform convention we assume is
\begin{equation}
	\tilde{h}(f)=\int_{-\infty}^{\infty}e^{i2\pi f t}h(t)dt.
\end{equation}
Equation \eqref{faith} calculates the overlap between two waveforms, with the same physical parameters, but maximized extrinsic (unphysical) parameters of little astrophysical interest, i.e. the time $t_c$ and phase $\Phi_c$ offsets. For the calculation of the inner product \eqref{inner product} that leads to Eq. \eqref{faith}, we have used realistic bounds of integration such that the lower and upper limits are $f_\text{min}=10^{-8}$ Hz and the Nyquist frequency $f_\text{max}=f_\text{Ny}$, respectively. Obviously, when comparing two equivalent signals we have $\mathcal{F}(h_1,h_1)=1$ and the inner product \eqref{inner product} satisfies the commutative law, thus $\mathcal{F}(h_1,h_2)=\mathcal{F}({h_2,h_1})$.

\subsection{Gravitational waves and Overlap}

In Fig. \ref{fig4.1} we plot some representative cases of GWs, detected by LISA, emitted by either geodesics around a Schwarzschild primary or around a primary surrounded by a dark matter halo. It is clear that the environment affects significantly the resulting waveforms, even when it is of galactic scale. At the early stage of the evolution, the vacuum and non-vacuum EMRI waveforms are in phase but only due to the fact that $7$ hours have elapsed. Nevertheless, after only $\sim 2$ months of observation, the GW signals dephase, with the dephasing becoming more significant as the compactness of the halo increases. Figure \ref{fig4.1} further demonstrates that the orbits inside halos dephase in a manner that designates that the GW frequencies should be redshifted, since the presence of dark matter leads to an increase in the geodesic's revolution period. After $\sim 7.5$ months of orbital evolution, all waveforms calculated have completely dephased, therefore the environment should play a very crucial role in EMRI evolution \cite{Cardoso:2021wlq,Cardoso:2022whc}.

Figure \ref{fig4.2} presents the faithfulness (maximized overlap) between vacuum Schwarzschild EMRIs and those with a primary residing in a halo. We have calculated the faithfulness of emitted GWs from geodesics around vacuum and non-vacuum primaries for varying compactness for the two and five months of observation (i.e. $t_\text{obs}=5\times 10^5 M_\text{BH}$ and $t_\text{obs}=1.3\times 10^6 M_\text{BH}$, respectively), which correspond roughly to $3000$ and $8000$ cycles for all halos considered. The choice of observation time seems to affect the faithfulness. Indeed, for shorter observation times the environment affects less the GW emission and propagation but longer observations lead generally to lower faithfulness. Similar analyses have been performed for even longer observation times, e.g. years \cite{Maselli:2020zgv,Liang:2022gdk,Barsanti:2022vvl}, though for circular equatorial EMRIs with small non-GR parameters. Nevertheless, our case is more sensitive to the generic, non-circular and precessing nature of orbits, as well as the strong effect the halo introduces to the secondary's trajectory when we increase it significantly.

When the compactness of the halo is arbitrarily small (of order $10^{-10}-10^{-7}$) the two waveforms differ by extremely little, if not at all, and the overlap is practically unity. Eventually it starts decreasing with the growth of the compactness of the halo since the GWs dephase significantly in the window of observation. Moreover, even when the compactness is of galactic scales, i.e. of order $10^{-6}-10^{-4}$, the overlap is still quite low and therefore EMRIs in a galactic environment can certainly be distinguishable from those in vacuum, especially when generic inspirals are considered. Of course, in our case, radiation reaction has not been taken into account, but our study does not put bounds on the initial conditions. Rather we let the orbits evolve in a generic manner by including off-equatorial, non-circular and precessing evolutions. This is the reason behind the rapid drop of the overlap; the signals become much more complicated, with multiply Fourier peaks, than the ones studied with radiation reaction but for equatorial and circular (see e.g. Refs. \cite{Cardoso:2021wlq,Maselli:2020zgv,Maselli:2021men,Barsanti:2022vvl,Zhang:2022hbt,Liang:2022gdk}) or eccentric EMRIs \cite{Barsanti:2022ana}. In fact, the claims in \cite{Barsanti:2022ana} that entertain the possibility of constraining better a non-GR scalar charge carried by the secondary, when the orbit is eccentric rather than circular, are in complete agreement with the qualitative picture of our findings, that is the more complicated the orbit, the more distinguishable are its EMRI parameters.

\subsection{Gravitational-wave frequency redshift}

Here, we GWs in the frequency domain in order to spot possible environmental effects in their spectra. Fig. \ref{fig5} depicts the Fourier harmonics of GWs with varying halo parameters. For reference, we include the frequencies of a vacuum EMRI with a Schwarzschild supermassive primary. As the compactness increases the GW frequencies are redshifted (shown in the left panel of Fig. \ref{fig5} and further observed earlier in Fig. \ref{fig4.1}). Physically, the redshift is associated with the presence of the halo which interacts with test particles, leads to dynamical friction and eventually increases their orbital period (as discussed in the previous subsection), thus decreasing the GW frequencies. We point out that there might be a case where the observability of the redshift could be potentially obscured by a change in the initial radial position of the geodesic or the mass of the primary. To the contrary, both changes will not only affect the amplitude of the GWs observed (and their respective Fourier amplitudes) but can also completely change the frequency domain spectrum if the orbit moves closer or further from the primary.

\subsection{Waveform imprints of compact environments}

Compact environments introduce a precession drift reversal, as shown in Sec. \ref{orb_prec}. Fig. \ref{fig6} focuses on the waveform frequencies of geodesics with varying initial radial position (same as Fig. \ref{fig3}) and halo parameters $M=3M_\text{BH}$, $a_0=10M$. Firstly, the Fourier peaks appear more concentrated to low frequencies, due to cumulative redshift, and exhibit a wealthier structure. For reference, the red, blue and green peaks correspond to trajectories with prograde, retrograde and no precession (see Fig. \ref{fig3}, \ref{fig4}). Precessing orbits consists of peak triplets, while the critical non-precessing geodesic possesses single Fourier peaks; an expected phenomenon due to the absence of precession frequencies. Each frequency triplet has a minimum and a maximum amplitude, with the position of the maximum (and the existence of the triplet) depending delicately to the initial position of the geodesic. Specifically, (retrograde-) prograde-precessing orbits, i.e. the ones lying (beyond) below the critical radius and the drop on the respective rotation curve, acquire maximal harmonics at the (leftmost) rightmost frequencies of each triplet, while the special non-precessing orbit has single harmonics which arise from the combination of the triplet.

In order to further elucidate the frequency evolution of EMRI systems in galactic halos and compact environments, i.e. the interchange between triplet maxima and minima, we can simulate an adiabatic `inspiral' through successive geodesics by simply changing the initial radial position of the geodesic and calculating the respective Fourier transforms of the resulting waveforms. We note that this is not the proper way of evolving inspirals, since we do not know how $E$ and $L_z$ evolve under radiation reaction, but the phenomenological imprint can be discerned since the fluxes should change dramatically slow in a timescale of $\sim 7.5$ months of evolution. 

In what follows, we keep $E/\mu=0.95$ and $L_z/\mu=3M_\text{BH}$ (even though for a realistic inspiral they should decrease in accord with the GW fluxes) and slowly change the initial condition $r(0)$ to obtain a geodesic evolution (similar to what is done for the Poincare maps). We then use the quadrupole formula to approximate the waveform and Fourier transform it to the frequency domain. From the Fourier peaks, we pick a particular frequency range which includes one harmonic and calculate the frequency for which the peak is maximized. We re-iterate the aforementioned method for different initial conditions to simulate a very rough estimate of the frequency evolution, though without knowing the actual phenomenological timescales involved.

Fig. \ref{fig7} depicts two cases of environments, namely a galactic-scale halo with $M=10^3 M_\text{BH}$, $a_0=10^5 M$ and a compact environment with $M=3M_\text{BH}$, $a_0=10 M$, where precession reversal occurs. The influence of the galactic environment in EMRI evolution is practically negligible and matches to good agreement that of vacuum EMRIs, i.e. as the radius of the secondary with respect to the primary decreases, the GW frequencies increase exponentially (see Fig. \ref{fig7}, left plot). 

The phenomenology is altered when the halo is ultracompact, as discussed in previous sections. Concentrated dark matter and the primary's influence on the test particle engage into a gravitational clash, which further can cancel out one another and lead to perfectly elliptic orbits that exhibit no precession (up to numerical precision). Fig. \ref{fig7} (right plot) shows the dependence of three subpeaks in a particular frequency regime with respect to the initial position (similar behavior is found for other harmonics). Firstly, we observe two subfrequencies (in red); one that grows in a similar manner as the one in the left plot of Fig. \ref{fig7}  and another that decays as the radius is decreased. These subfrequencies correspond to the leftmost (solid red curve) and the rightmost (dotted-dashed) subpeaks of Fig. \ref{fig6}. The central subfrequency of the orbit remains constant, which is expected since the orbit, precessing or counter-precessing, still contains the revolution frequency of the elliptic orbit. Interestingly, all subpeaks meet at a critical radius (black dashed line in Fig. \ref{fig7}), to form a single peak, that coincides
with the frequency of the critical non-precessing geodesic. 

Unfortunately, the absence of radiation reaction effects forbid us to simulate a proper inspiral, therefore we cannot make any solid predictions on the timescales involved when all subpeaks merge and how long the secondary can latch in such a special orbit, as well as how discernible this effect may be with future space-based detectors. In any case, we have shown that dark matter environments (compact or not) affect GW generation and propagation in various ways, and at significant levels, and have the potential to introduce direct phenomenological imprints (see also \cite{Cardoso:2022whc}) that can serve as further `smoking guns' of dense dark matter clumps around supermassive BHs.

\section{Discussion}

We have investigated the phase space of geodesics of a newly-obtained exact solution of GR that describes a Schwarzschild BH surrounded by a dark matter halo which one can tune its compactness. The orbits on such geometry, together with the characteristic orbital frequencies, behave in a similar manner as those around vacuum Schwarzschild BHs when the compactness is tuned to astrophysical values, i.e. $M/a_0\lesssim 10^{-5}$, that describe galactic dark matter halos. Nevertheless, the compactness can be further increased to simulate BH hair and dense environments. This is when the geodesics experience significant effects. The available phase space volume is enlarged with respect to that of Schwarzschild, that leads to a redshift in the rotation numbers of geodesics and their respective orbital resonances.

Interestingly, when the dark matter is concentrated around the central BH, a delicate gravitational competition takes place between general-relativistic effects and the dark matter influence. In these cases, we have found critical radii for which the rotation curves reach unity and then diminish to zero. These critical points designate a transition from prograde to retrograde precession drift, and exactly at these points in phase space the corresponding geodesic experiences no precession. Similar results have been obtained recently in dense dark matter cores \cite{Igata:2022nkt,Igata:2022rcm} which further justifies the validity of the orbital analysis presented here.

We have further analyzed the dominant GWs emitted from these geodesics in an attempt to visualize potential phenomenological imprints sourced by EMRIs, where the primary of the binary is described by the galactic BH model. Under the assumption that the space-based detector is LISA, we have found that increasing the compactness of the halo leads to a rapid dephasing of GWs due to redshift; an outcome attributed to the presence of the dark matter field that leads to dynamical friction and to the growth of the secondary's orbital period. In fact, since the orbits we study here are not bound to be circular or equatorial, but rather generic, give rise to a quick drop in faithfulness between GWs in vacuum and non-vacuum binaries as $M/a_0$ grows from galactic-scales to compact BH environments. The redshift of rotation numbers at the orbital level translates to a typical GW frequency redshift during `inspiral'. Similar redshift has been found in the quasinormal modes of the remnant after merger \cite{Cardoso:2021wlq}, while more intricate phenomena occur when one takes into account both axial and polar GW fluxes \cite{Cardoso:2022whc}.

Nevertheless, the waveform spectra resulting from binaries surrounded by very compact environments tell a different story. Each Fourier harmonic breaks onto three subpeaks, where the two outmost ones are interchanged during the retrograde-to-prograde transition, while the central one remains constant. We have simulated a very rough EMRI evolution through the turning point by utilizing consecutive geodesics with decreasing initial conditions (though the energy and angular momentum of the secondary are kept constant) and found that right at the point of no precession all three subpeaks combine into the central harmonic peak. Future space-based detectors such as LISA should, therefore, be able to discern the existence of both galactic-scale and dense environment surrounding EMRIs, through cumulative dephasing and retrograde-to-prograde precession drifts, thus environmental effects should be taken into consideration when building accurate EMRI waveforms.

Although in this paper we explored environmental effects in EMRIs at the geodesic level, proper inspirals should be driven by GW fluxes, and more precisely by gravitational self-force effects \cite{Barack:2009ux}. The axial and polar fluxes due to metric perturbations introduced by the secondary have only been calculated for circular, equatorial EMRIs in astrophysical environments very recently \cite{Cardoso:2021wlq,Cardoso:2022whc}. A faithful direction which we are currently pursuing is to accurately evolve an EMRI surrounded by a compact dark matter cloud, in order to examine the exact timescales involved in retrograde-to-prograde orbital transitions and understand if these effect can be distinguished with space-borne interferometers. Another direction to explore is the connection of the galactic BH solution with the recently found spectral instabilities that quasinormal modes suffer from in BH physics \cite{Nollert:1996rf,Nollert:1998ys,Daghigh:2020jyk,Jaramillo:2020tuu,Destounis:2021lum,Jaramillo:2021tmt,Jaramillo:2022kuv,Cheung:2021bol,Berti:2022hwx,Boyanov:2022ark,Yang:2022wlm,Konoplya:2022hll,Konoplya:2022pbc}. Since it has been shown that the presence of an astrophysical environment sources fluid modes \cite{Cardoso:2022whc} that couple with polar metric perturbations and `destabilize' the quasinormal mode spectrum it is worth using non-modal tools, such as the pseudospectrum \cite{Trefethen:2005}, to further elucidate the existence of such modes and their effect in astrophysical BH settings including realistic environments.

\begin{acknowledgments}
	The authors would like to warmly thank Theocharis Apostolatos, Vitor Cardoso, Francisco Duque, Rodrigo Panosso Macedo and Andrea Maselli for helpful discussions. K.D. acknowledges financial support provided under the European Union's H2020 ERC, Starting Grant agreement no.~DarkGRA--757480 and the MIUR PRIN and FARE programmes (GW-NEXT, CUP: B84I20000100001). This work was supported by the DAAD program for the ``promotion of the exchange and scientific cooperation between Greece and Germany IKYDAAD 2022" (57628320).
\end{acknowledgments}

\appendix

\section{Integrability of geodesics in spherically-symmetric geometries}\label{appA}

Let $(\mathcal{M},\mathbf{g})$ be a four-dimensional, Haussdorf space, with Lorentzian signature and no boundary, such that there is a local coordinate system ${x}^{\alpha}=(t,r,x,\phi)$ -- part of the atlas -- in which the line element
assumes the form
\begin{equation}
	ds^2=f_{1}(r)dt^{2}-f_{2}(r)dr^{2}-r^{2}\Big(\frac{1}{(1-x^{2})}dx^{2}+(1-x^{2})d\phi^{2}\Big).
\end{equation}
Note that if $x$ is to be identified as $\cos\theta$ then the usual standard Schwarzschild-like coordinates are recovered. 

The metric tensor field under consideration is susceptible to four linearly-independent (with constant coefficients) KVFs, which in the local coordinate system assume the form
\begin{align}
	&\eta^{\alpha}=(1,0,0,0),\\
	&\xi_{(1)}^{\alpha}=(0,0,\sqrt{1-x^{2}}\cos\phi,x\frac{\sin\phi}{\sqrt{1-x^{2}}}),\\
	&\xi_{(2)}^{\alpha}=(0,0,\sqrt{1-x^{2}}\sin\phi,-x\frac{\cos\phi}{\sqrt{1-x^{2}}}),\\
	&\xi_{(3)}^{\alpha}=(0,0,0,1),\\
	&\pounds_{\ainF{\eta}}g_{\alpha\beta}=0,\\
	&\pounds_{{\ainF{\xi}}_{(i)}}g_{\alpha\beta}=0,\, i\in\{1,2,3\},
\end{align}
where $\pounds_{\ainF{X}}$ is the Lie derivative of the vector field $\ainF{X}$. The corresponding (closed) Lie algebra is
\begin{align}
	&[\ainF{\eta},\ainF{\xi}_{(i)}]=0,\\
	&[\ainF{\xi}_{(i)},\ainF{\xi}_{(j)}]=\ainF{\xi}_{(k)}\epsilon^{k}_{\phantom{1}ij},\, i,j,k\in\{1,2,3\},
\end{align}
where $\epsilon^{k}_{\phantom{1}ij}$ is the Levi-Civita symbol in three dimensions with Euclidean signature. The first KVF acts simply transitively, while the last three multiple transitively. The important problem here is to solve the geodesics equations \eqref{geodesic}. It is a trivial exercise to show that if 
\begin{equation}
\pounds_{\ainF{X}}g_{\alpha\beta}=0,
\end{equation}
i.e., if the vector field \ain{X} is a KVF then the quantity 
\begin{equation}
\mathcal{I}\equiv X^{\alpha}g_{\alpha\beta}\dot x^{\beta}
\end{equation}
is an integral of motion for the geodesics equation, that is
\begin{equation}
\dot x^{\alpha}\nabla_{\alpha}\mathcal{I}=0.
\end{equation}
Therefore in the present case there are four integrals of motion. Nevertheless, due to the multiply-transitive character of a part of the entire symmetry group, along with the manifest appearance of the trigonometric functions, we can easily deduced that if
\begin{align}
	&\mathcal{I}_{1}\equiv\eta^{\alpha}g_{\alpha\beta}\dot x^{\beta}, \,\,\,\mathcal{I}_{(i)}\equiv\xi_{(i)}^{\alpha}g_{\alpha\beta}\dot x^{\beta},
\end{align}
then the system
\begin{align}
	&\mathcal{I}_{1}=E,\\
	&\mathcal{I}_{(1)}^{2}+\mathcal{I}_{(2)}^{2}+\mathcal{I}_{(3)}^{2}=L^{2},\\
	&\mathcal{I}_{(3)}=-L_{z},
\end{align}
can be solved in terms of the generalized velocities $\dot t, \dot x, \dot\phi$. Then, substitution of those relations (and their derivatives with respect to the affine parameter $\tau$) into Eq. \eqref{geodesic} results in only one component equation (the other three being empty)
\begin{equation}
\ddot r+\frac{\partial_{r}f_{2}(r)}{2f_{2}(r)}\dot r^{2}
+\frac{E^{2}\partial_{r}f_{1}(r)}{2f_{1}(r)^{2}f_{2}(r)}
-\frac{L^{2}}{f_{2}(r)^{3}}=0,
\end{equation}
where $r=r(\tau)$. Obviously, special forms for the functions $f_{1}, f_{2}$ may lead to further symmetries via the Lie-point (or contact or even dynamical) symmetries of the last equation.

\bibliography{halo}

\end{document}